 \documentclass[sigconf]{acmart}
\AtBeginDocument{%
	\providecommand\BibTeX{{%
			\normalfont B\kern-0.5em{\scshape i\kern-0.25em b}\kern-0.8em\TeX}}}

\acmConference[KDD 2023]{}{August 6, 2023}{Long Beach, California, United States}

\usepackage{ mathdots  }
\usepackage{ amsmath   }
\usepackage{ amssymb   }
\usepackage{ color     }
\usepackage{ graphicx  }
\usepackage{ amsfonts  }
\usepackage{ epstopdf  }
\usepackage{ upgreek   }
\usepackage{ bm        }
\usepackage{ graphicx  }
\usepackage{ subfigure }
\usepackage{ mathtools }
\usepackage{ algorithm }
\usepackage{algorithmic}
\usepackage{ booktabs  }
\usepackage{ multirow  }
\usepackage{ caption   }
\usepackage{ array     }
\usepackage{ float     }
\usepackage{   soul    }
\usepackage{enumitem}
\usepackage{multirow}
\usepackage[normalem]{ulem}
\newcommand{\norm}[1]{\left\lVert#1\right\rVert}
\usepackage{bbm}
\useunder{\uline}{\ul}{}

\DeclareMathOperator{\argmax}{arg\,max}

\begin{document}
	% ===========================
	%          title
	% ===========================
	\title{Multi-factor Sequential Re-ranking with Perception-Aware Diversification}
	
	% ===========================
	%          author
	% ===========================
	\author{Yue Xu}
	\email{yuexu.xy@foxmail.com}
	\affiliation{%
	\institution{Alibaba Group.}
	}
	
	\author{Hao Chen}
	\email{sundaychenhao@gmail.com}
	\affiliation{%
	\institution{The Hong Kong Polytechnic University, Hong Kong.}
	}
	\author{Zefan Wang}
	\email{wongzfn@gmail.com}
	\affiliation{%
	\institution{Jinan University, China.}
	}
	\author{Jianwen Yin}
	\email{yjw264077@alibaba-inc.com}
	\affiliation{%
	\institution{Alibaba Group.}
	}
	
	\author{Qijie Shen}
	\email{qijie.sqj@alibaba-inc.com}
	\affiliation{%
	\institution{Alibaba Group.}
	}
	
	\author{Dimin Wang}
	\email{dimin.wdm@alibaba-inc.com}
	\affiliation{%
	\institution{Alibaba Group.}
	}
	
	\author{Feiran Huang}
	\email{huangfr@jnu.edu.cn}
	\affiliation{%
	\institution{Jinan University, China.}
	}
	
	\author{Lixiang Lai}
	\email{lixiang.llx@alibaba-inc.com}
	\affiliation{%
	\institution{Alibaba Group.}
	}
	
	\author{Tao Zhuang}
	\email{zhuangtao.zt@alibaba-inc.com}
	\affiliation{%
	\institution{Alibaba Group.}
	}
	
	\author{Junfeng Ge}
	\email{beili.gjf@alibaba-inc.com}
	\affiliation{%
	\institution{Alibaba Group.}
	}
	
	\author{Xia Hu}
	\email{xia.hu@rice.edu}
	\affiliation{%
	\institution{Rice University, USA}
	}
    % format 1
%    \author{\centerline{Yue Xu$^1$, Hao Chen$^2$, Zefan Wang$^3$, Jianwen Yin$^1$, Qijie Shen$^1$, Dimin Wang$^1$}}
%    \author{\centerline{Feiran Huang$^3$, Lixiang Lai$^1$, Tao Zhuang$^1$, Junfeng Ge$^1$, Xia Hu$^4$}}
%    \affiliation{%
%        \institution{\centerline{$^1$Alibaba Group, China \hspace{0.3em} $^2$The Hong Kong Polytechnic University, Hong Kong}}
%    }
%    \affiliation{%
%        \institution{\centerline{$^3$Jinan University, China \hspace{0.3em} $^4$Rice University, USA}}
%    }
%    \affiliation{
%        \institution{\centerline{\{guyue.yuexu, yjw264077, qijie.sqj, dimin.wdm, lixiang.llx, zhuangtao.zt, beili.gjf\}@alibaba-inc.com}}
%    }
%    \affiliation{
%        \institution{\centerline{sundaychenhao@gmail.com, wongzfn@gmail.com, huangfr@jnu.edu.cn, xia.hu@rice.edu}}
%    }

	% ============================
	%          abstract
	% ============================
	\begin{abstract}
    Feed recommendation systems, which recommend a sequence of items for users to browse and interact with, have gained significant popularity in practical applications. In feed products, users tend to browse a large number of items in succession, so the previously viewed items have a significant impact on users' behavior towards the following items. Therefore, traditional methods that mainly focus on improving the accuracy of recommended items are suboptimal for feed recommendations because they may recommend highly similar items. For feed recommendation, it is crucial to consider both the accuracy and diversity of the recommended item sequences in order to satisfy users' evolving interest when consecutively viewing items. To this end, this work proposes a general re-ranking framework named Multi-factor Sequential Re-ranking with Perception-Aware Diversification~(MPAD) to jointly optimize accuracy and diversity for feed recommendation in a sequential manner.
    Specifically, MPAD first extracts users' different scales of interests from their behavior sequences through graph clustering-based aggregations. Then, MPAD proposes two sub-models to respectively evaluate the accuracy and diversity of a given item by capturing  users' evolving interest due to the ever-changing context and users' personal perception of diversity from an item sequence perspective. This is consistent with the browsing nature of the feed scenario. Finally, MPAD generates the return list by sequentially selecting optimal items from the candidate set to maximize the joint benefits of accuracy and diversity of the entire list. MPAD has been implemented in Taobao's homepage feed to serve the main traffic and provide services to recommend billions of items to hundreds of millions of users every day.
	\end{abstract}
	
	% ============================
	%        Keywords
	% ============================
	\maketitle
	\pagestyle{plain} % removes running headers
	
% ============================
%        Introduction
% ============================
\section{Introduction}
    \begin{figure}[t]
    \centering
    \subfigure[Channel blocks]
    { 	
    	\label{fig:channelRec}
    	\includegraphics[trim = 10 10 10 10, clip, width=0.35\columnwidth]{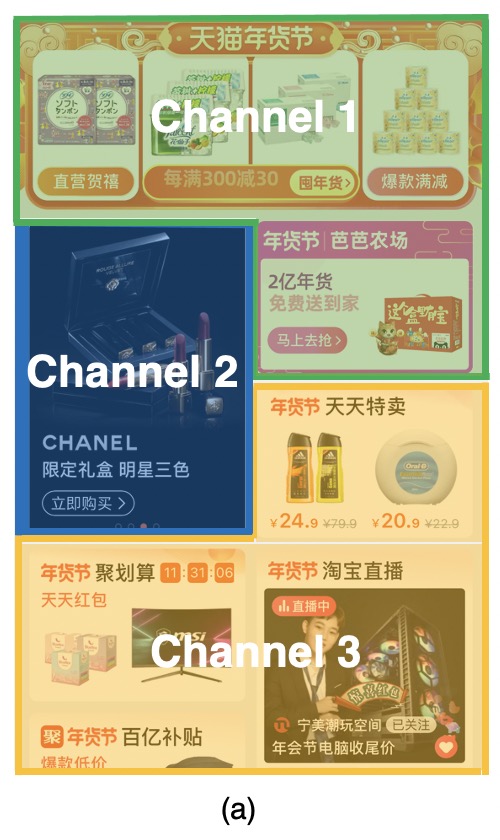}
    }
    \subfigure[Feed recommendation]
    { 	
    	\label{fig:feedRec}
    	\includegraphics[trim = 10 10 10 10, clip, width=0.35\columnwidth]{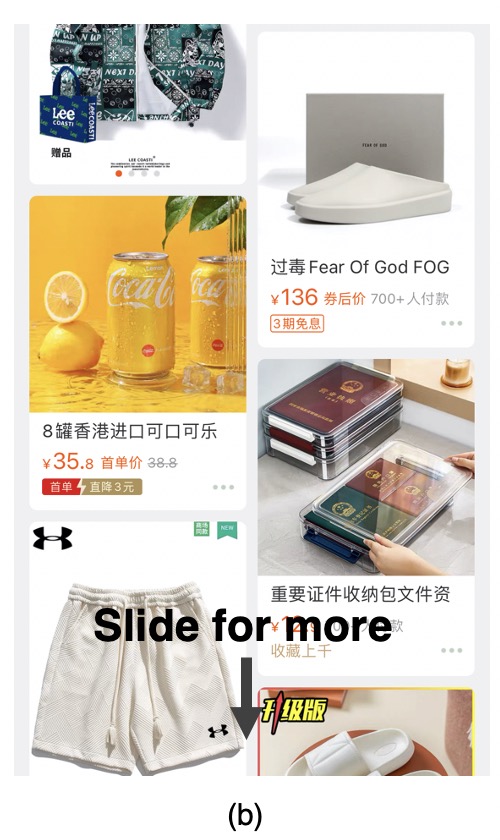}
    }
    \caption{An example of different forms of real-world RS. Left: Traditional RS recommends items in different channel blocks. Right: Feed RS recommends a sequence of items where users can slide down to view more items.}
    \label{fig:feedExample}
    \end{figure}

    Feed recommendation system~(RS) is one type of product that recommends a sequence of items for users to browse and interact with. 
    It has been widely applied to various online platforms, such as on the homepage of Kuaishou~\cite{lin2022feature}, Xiaohongshu~\cite{huang2021sliding}, Taobao~\cite{qian2022intelligent}, and AliExpress~\cite{hao2021re}. 
    An example of the feed recommendation is given in Figure~\ref{fig:feedExample}.
    The feed allows users to continuously scroll down the item list of items in the viewing window, such that previously viewed items have a large impact on users' behaviors towards the next item. 
    In this case, traditional methods that mainly focus on improving the accuracy of recommended items become sub-optimal for feed recommendation because they usually ignore the correlations between consecutive items. For example, if a user was shown a mobile phone item, it may be sub-optimal to put a series of more mobile phone items next to it. This mismatch is exacerbated by the fact that similar items tend to have similar click ratios.
    Therefore, it is of vital importance for feed recommendation methods to consider both accuracy and diversity from an item sequence perspective to attract users to browse and interact with more items in the feed~\cite{lin2022feature,huang2021sliding,qian2022intelligent}.
    
    % Perception-aware diversity
    This work focuses on two common characteristics that are closely related to optimizing accuracy and diversity in feed recommendations.
    First, different users may have different perceptions of diversity, such that item diversity in feed recommendations should be measured based on their personal interests.
    Existing works on diversified recommendations mainly focus on measuring the dissimilarity between item pairs, without considering  users' personal interests. 
    For example, the post-processing methods improve diversity by heuristically rearranging the item order based on predefined rules, which are not customized for all users~\cite{ashkan2015optimal,boim2011diversification,borodin2012max,carbonell1998use,qin2013promoting,sha2016framework}; the learning-based methods measure the similarity of a given item pair by directly comparing the item embeddings~\cite{abdool2020managing,zheng2021dgcn,wilhelm2018practical,chen2018fast}.
    Though effective, the ignorance of users' personal interests may lead to a mismatch between the model's definition of diversity and users' perceptions of diversity. 
    For example, some female customers prefer to view more clothing than others in the feed. 
    Directly reducing the probability of presenting user-preferred items to increase diversity may degrade their satisfaction with the recommended results.

    Second, in feed applications, users tend to view many items in a row, such that users' interests may evolve during this continuous browsing process. In this case, the measurement of both item accuracy and diversity should consider the evolving interest due to the ever-changing context so as to accommodate the sequential browsing nature in feed scenarios. 
    However, most existing interest models mainly focus on learning users' interests from their historical behaviors with less emphasis on the evolution of interests along with the browsing context~\cite{zhou2018deep,zhou2019deep,qi2020search,chen2022efficient}. 
    Another line of research on re-ranking proposes various list-wise solutions to capture the interior correlations between items within the context~\cite{ai2018learning,pei2019personalized,bello2018seq2slate,huang2020personalized}. Nevertheless, they mainly focus on improving accuracy regardless of diversity. 
    Some recent works devote efforts to solve this accuracy-diversity dilemma and obtained promising results~\cite{wilhelm2018practical,zheng2021dgcn,huang2021sliding,lin2022feature}. However, the joint optimization of accuracy and diversity still remains to be a challenging problem, especially for industrial implementation on large-scale systems.

    In light of the above challenges, in this paper, we investigate the following research questions. 
    1)~How to formulate and design a general framework for jointly optimizing accuracy and diversity from an item sequence perspective? 
    2)~How to estimate accuracy and diversity with adaptation to the evolution of user interest while browsing consecutive items in feed scenarios?
    3)~How to implement the proposed framework in industrial systems for practical applications and how well does it perform?
    To this end, we propose a general Multi-factor Sequential Re-ranking with Perception-Aware Diversification~(MPAD) framework to jointly optimize accuracy and diversity for practical feed recommendations. 
    This framework consists of four main components.
    The bi-sequential determinantal point process~(BS-DPP) algorithm provides a principled and tractable framework for sequential item selection to maximize the joint benefits of accuracy and diversity of the entire item list. 
    The Multi-scale Interest Extraction~(MIE) model extracts multi-scale user interests through graph clustering-based aggregations. 
    The Context-aware Accuracy Estimation~(CAE) model provides an estimate of context-aware accuracy from a sequence perspective by learning from both the multi-scale interests and the ever-changing browsing context.
    The Perception-Aware Kernel~(PDK) evaluates the similarity between items with consideration of the user's perception of diversity based on personal interests. 
    The main contributions are as follows.
    \begin{itemize}[leftmargin=*]
    \item This work formulates the feed recommendation task as a multi-factor re-ranking problem and proposes a principle and tractable MPAD framework to maximize the joint benefits of accuracy and diversity of the entire recommended item list.
    \item This work proposes a series of collaborative models to estimate the accuracy and diversity of an item list from a sequence perspective. They are able to capture the influence from both the browsing context and the evolving user interests to align with the browsing nature of the feed scenario. We also propose a tailored BS-DPP algorithm to jointly optimize the accuracy and diversity when selecting optimal items in a sequential manner.
    \item This paper presents a general system architecture for the deployment of MPAD in industrial systems. It has now been implemented in the homepage feed to achieve $ 2.4\% $ lift on user clicks, $ 2.0\% $ lift on stay time, and $ 4.0\% $ lift on content diversity. The architecture now serves Taobao's main traffic with $ 120,000 $ queries-per-second at peak. 
    \item This work conducts extensive experiments on both offline datasets and online A/B tests. The results show that our proposed MPAD significantly outperforms other methods. The source code has been made public\footnote{The source code is available at https://anonymous.4open.science/r/MPAD/.}.
    \end{itemize}
	
	% ============================
	%          Section
	% ============================
	\section{Problem Setup}
%	Figure~\ref{fig:structure} demonstrates a typical pipeline of industrial RS~\cite{wilhelm2018practical,huang2021sliding}. 
	A typical pipeline of industrial RS includes three stages~\cite{wilhelm2018practical,huang2021sliding}, i.e., matching, ranking, and re-ranking.
	The RS first retrieves candidate items from item databases at the matching stage. Then, the ranking modules measure item accuracy in a point-wise manner. Finally, the top items will be sent to the re-ranking module to determine the final item list to present to users.

	In this paper, we consider a multi-factor feed recommendation problem at the re-ranking stage, where the task is to select an item sequence $ S = \{i_1,i_2,\cdots,i_K\} $ with size $ K $ from a candidate set $ I = \{i_1, i_2, \cdots, i_N\} $ with size $ N \gg K $ provided by the ranking module. 
	The selection of sequence $ S $ depends on both the \textit{item accuracy} which relates to user's preference for the items and the list-wise \textit{item diversity} which influences user's intention to browse and interact in practical applications~\cite{cheng2017learning,wilhelm2018practical,huang2021sliding,liu2022neural}.
	Formally, given a target user $ u $ and a set of candidate items $ I $, our aim is to select a fixed-size subset from $ I $ and determine their order in a page to maximize a joint utility function: 
	\begin{align}\label{P0}
	\mathcal{P}_0: \arg\max_{S \subseteq I} f(u, S) = F(\text{Acc}(u, S), \text{Div}(u, S)),
	\end{align}
	where the first term $ \text{Acc}(u, S) $ evaluates the context-aware item accuracy based on user interests and browsing context, the second term $ \text{Div}(u, S) $ evaluates the list-wise diversity of all items, and the fusion function $ F(\cdot) $ measures the contribution of item accuracy and diversity to the joint utility $ f(u, S) $. 
	
	Note that this formulation extends the commonly used item-level diversity~\cite{lin2022feature,wilhelm2018practical,huang2021sliding,chen2018fast} to personalized user-item-level diversity, i.e., evolving from $ \text{Div}(S) $ to $ \text{Div}(u, S) $. As such, the solution needs to consider user's \textit{personalized} perception of diversity on the recommended results. 

	\section{Methodology}
	\begin{figure*}[t]
		\centering
		\includegraphics[trim = 0 0 0 0, clip, width=2\columnwidth]{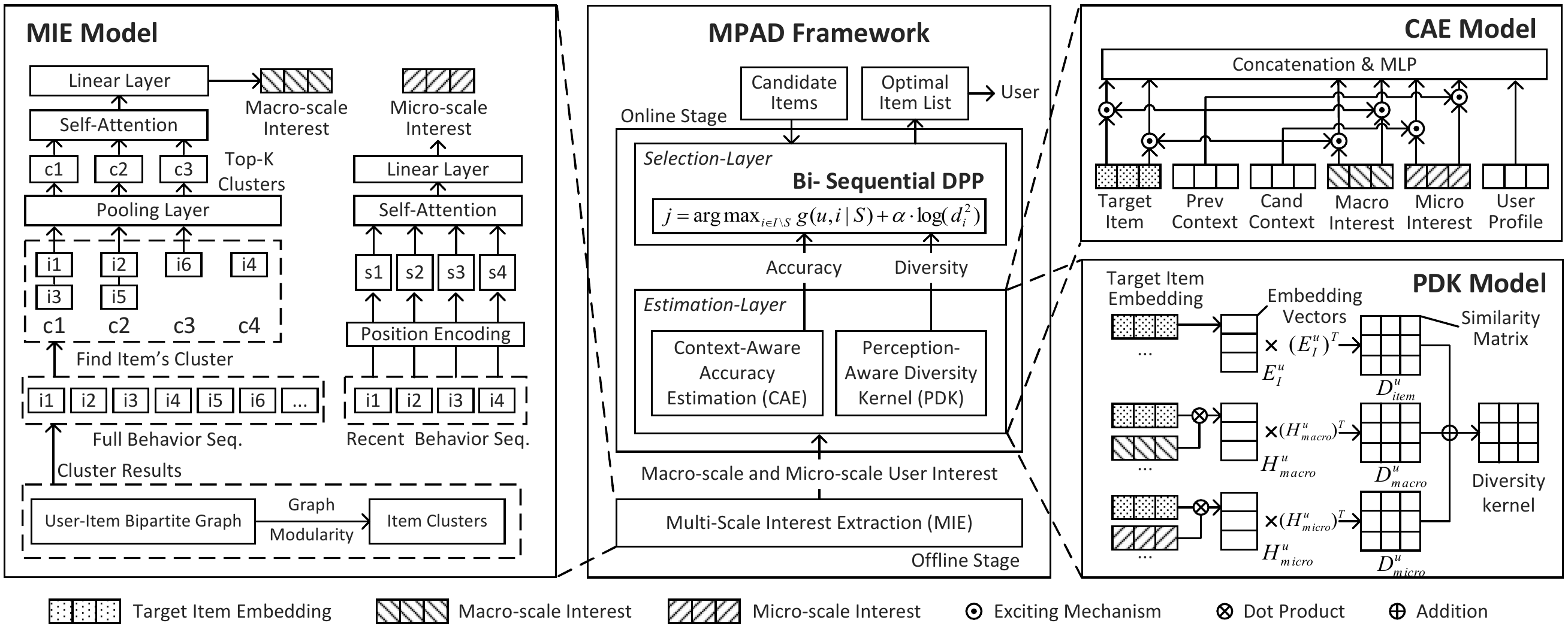}
		\caption{An overview of the MPAD framework.}
		\label{fig:MPAD}
	\end{figure*}
	This section first gives an overview of the proposed MPAD framework in Sec.~\ref{sec:framework}. 
	Then, this section introduces the main building blocks of MPAD in order from Sec.~\ref{sec:BS-DPP} to Sec.~\ref{sec:PDK}.
	Finally, this section discusses the online implementation of MPAD in Sec.~\ref{sec:implementation}.
	
	\subsection{Framework Overview}
	\label{sec:framework}
	The framework consists of two layers: the \textit{selection-layer} uses a sequential item selection algorithm to select items from the candidate set using the item accuracy and diversity scores evaluated by the \textit{estimation-layer}.
	The detailed workflow is presented in Figure~\ref{fig:MPAD}.
	
	Specifically, the selection-layer operates with the BS-DPP algorithm which considers both list-wise item diversity and context-aware item accuracy during the selection of optimal items. It indeed offers a principle and tractable solution for function $ F(\cdot) $. 
	The estimation-layer, on the other hand, consists of three components.
	The \textit{first} component is MIE which groups users and items into different clusters and represents the user's multi-scale interest based on behavior sequences encoded by item/cluster embeddings. MIE can be computed offline for online complexity reduction.
	The \textit{second} component is CAE which refines the point-wise accuracy scores from the ranking stage into context-aware accuracy scores by making use of both browsing context and multi-scale user interests. 
	The refined scores are used in the computation of $ \text{Acc}(u, S) $ in BS-DPP. 
	The \textit{third} component is PDK which computes item similarities based on both item embedding and the user's different scales of interests. The diversity kernel measures the diversity term $ \text{Div}(u,S) $ in BS-DPP.
	
	\subsection{Bi-Sequential Item Selection}
	\label{sec:BS-DPP}
	This section presents the BS-DPP algorithm for incremental item selection at the selection-layer. 
	In BS-DPP, both the item diversity scores and the item accuracy scores are considered to be sequentially updated along with the item selection process.
	
	This is different with standard Determinantal Point Process~(DPP) methods~\cite{chen2018fast,wilhelm2018practical} where the item accuracy scores are considered to be fixed values, regardless of the change of context. 
	Therefore, BS-DPP is more in accordance with the browsing nature of feed products where the user's interests may evolve during reviewing consecutive items.  
	
	\subsubsection{Task Formulation}
	A point process $ P $ defined on an item set $ I = \{i_1,i_2,\cdots,i_N\} $ is a probability distribution on the powerset of $ I $~(i.e., the set of all subsets of $ I $), where the probability satisfies $ \sum_{S \subseteq I} P(S) = 1 $.  
	The probability of choosing a specific item subset is determined by the kernel function in DPP and the item selection process is usually modeled as a MAP inference~\cite{kulesza2012determinantal,chen2018fast}. 
	In this paper, we define the DPP kernel based on a combined measurement of $ \text{Acc}(u,S) $ and $ \text{Div}(u,S) $, such that the probability of choosing an item subset is naturally proportional to the joint optimization of item accuracy and diversity.
	Based on the DPP theory~\cite{kulesza2012determinantal,chen2018fast}, the objective in~\eqref{P0} equals to:
	\begin{equation}\label{P1}
		\mathcal{P}_1: \arg \max_{S \in I} F(\text{Acc}(u, S), \text{Div}(u, S)) = \log\det(\boldsymbol{K}^u_{S}),
	\end{equation}
	where $ \boldsymbol{K}^u_{S} $ is the kernel function defined with $ \text{Acc}(u,S) $ and $ \text{Div}(u,S) $, to be discussed later; $ \log\det(\boldsymbol{K}^u_{S}) $ is the log-probability function of choosing a subset $ S $ for user $ u $.
	In this way, the aim to maximize the utility function $ f(u, S) $ in~\eqref{P0} is transformed into maximizing the log-probability function $ h(u, S) = \log\det(\boldsymbol{K}^u_{S})$.
	
	\subsubsection{Bi-Sequential DPP}
	Standard DPP methods~\cite{chen2018fast,wilhelm2018practical} construct the kernel matrix as follows:
	\begin{equation}\label{eq:originKernelFunc}
		K^u_S(i,j) = g(u,i)\cdot D(i,j) \cdot g(u,j),
	\end{equation}
	where $ g(u,i) $ is the point-wise accuracy score evaluated between user $ u $ and item $ i \in S $, regardless of the page-context, while $ D(i,j) $ measures the similarity between item $ i $ and item $ j $ with $ \forall i,j \in S $, regardless of user's personal interests.
	In contrast, BS-DPP considers that 1)~the accuracy scores are related to the browsing context, i.e., the previously added items in $ S $; and 2) the diversity scores are related to the user's interests. 
	This changes the definition in~\eqref{eq:originKernelFunc} into
	\begin{equation}\label{eq:kernel}
		K^u_S(i,j) = g(u,i|S)\cdot D(i,j|E_u) \cdot g(u,j|S),
	\end{equation}
	where $ g(u,i|S) $ denotes the context-aware accuracy score which conditions on the previously presented items in $ S $, while $ D(i,j|E_u) $ measures the similarity between item $ i,j \in S $ condition on the user's interest $ E_u $.
	We modify the log-probability of choosing a subset $ S $ as
	\begin{equation}\label{eq:logProbability}
		h(u, S) = \sum\nolimits_{i\in S} g(u,i|S) + \alpha \cdot \log\det(\boldsymbol{D}^u_{S}),
	\end{equation}
	where $ \alpha $ is a tunable parameter to control the trade-off between the diversity and accuracy of the recommended results.
	It is useful in practical feed applications since different platforms need such a parameter to control the tendency towards accuracy or diversity to suit different business orientations, e.g., more accuracy for relevant recommendations or more diversity for discovering new interests.
	
	The objective in~\eqref{P1} can be solved based on the popular greedy approximation methods~\cite{chen2018fast,wilhelm2018practical,gartrell2018deep}, which maximize the marginal gain when incrementally adding a new item to set $ S $. 
	Combining with our definition of the kernel function, the greedy maximization step to choose an optimal item per iteration can be written as
	\begin{subequations}\label{eq:greedyMaximization}
		\begin{align}
			j &= \argmax_{i \in I \setminus S} \log\det\left(\boldsymbol{K}^u_{S \cup \{i\}}\right) - \log\det\left(\boldsymbol{K}^u_{S}\right) \\
			& =\argmax_{i \in I \setminus S} h\left(u, S \cup \{i\}\right) - h(u, S) \\
			&= \arg\max_{i \in I \setminus S} g(u,i|S) \!+\! \alpha \cdot \left(\log \det \left(\boldsymbol{D}^u_{S\cup\{i\}}\right) \!-\! \log\det\left(\boldsymbol{D}^u_{S}\right)\right) \label{opt1-2} \\
			&= \arg\max_{i \in I \setminus S} g(u,i|S) + \alpha \cdot \log (d^2_i). \label{eq:finalSelectionStep}
		\end{align}
	\end{subequations}
	
	The complete algorithm of BS-DPP in MPAD is given in Algorithm~\ref{algorithm}. We defer more details on the derivations of~\eqref{eq:greedyMaximization} and the update of term $ \log(d_i^2) $, i.e., Step.~6 and Step.~7 in Algorithm~\ref{algorithm}, to the appendix.
	We now introduce how to obtain $ g(u,i|S) $ and $ \log (d^2_i) $ in~\eqref{eq:finalSelectionStep} via the proposed CAE and PDK model in the sequel. 
	\floatname{algorithm}{Algorithm} 
	\begin{algorithm}[t]
		\caption{Bi-Sequential Item Selection in MPAD}
		\label{algorithm}
		\begin{algorithmic}[1]
			\STATE \textbf{Initialization:} 
			\STATE{$\boldsymbol{D}_S$, $ \epsilon $, ${\bf c}_i=[]$, $d_i^2=\boldsymbol{D}_{ii}$, $j=\argmax_{i\in{I}}\log(d_i^2)$, $S=\{j\}$.}
			\STATE \textbf{Iteration:}
			\WHILE{$ |S|<k $ and $ d^2_i < \varepsilon $}
			\FOR{$ i \in I \setminus S $}
			\STATE $e_i=(\boldsymbol{D}_{ji}-\langle{\bf c}_j,{\bf c}_i\rangle)/d_j$.
			\STATE Update $d_i^2=d_i^2-e_i^2$, ${\bf c}_i=[{\bf c}_i \quad e_i]$.
			\STATE Update $ g(u,i|S) $ with the proposed preference model. 
			\ENDFOR
			\STATE Obtain $j = \arg\max_{i \in I \setminus S} g(u,i|S) + \alpha \cdot \log (d^2_i)$.
			\STATE Update subset $ S = S \cup \{j\} $.
			\ENDWHILE
		\end{algorithmic}
	\end{algorithm}

	\subsection{Multi-Scale Interest Extraction}
	\label{sec:interest}
	In this section, we propose the MIE model to extract users' multi-scale interests, which are used as the input of the subsequent CAE and PDK models.
	Existing user interest models usually directly perform self-attention on user's behavior items~\cite{zhou2018deep,pi2019practice,qi2020search,chen2022efficient}. However, directly mixing the information from a large quantity of raw item-level features may introduce redundant or noisy information to the model thus affecting learning performance. It is also hard for them to distinguish users’ different aspects of interests, especially from the full behavior sequences.
	
	% Therefore, this work proposes MIE to describe user interests from two scales, i.e., time-scale such as long-term and short-term interests, and aspect-scale such as different interest points on clothing, dressing, or sporting.
 	Therefore, this work proposes MIE to describe user interests from two scales, i.e., the micro-scale and the macro-scale. The micro-scale captures users' recent interests, such as their recent attention to gold necklaces and earrings. The macro-scale, on the other hand, models the user's long-term interests at a broader scope, such as fashion, clothing, or sports. For macro-scale interests, MIE groups items into clusters based on graph modularity and represents the user's macro-level interest through cluster-wise aggregated embeddings. Each cluster corresponds to one interest point at the macro level. 
    For micro-scale interests, MIE directly uses item-level features of each item within the user's behavior sequence, such as item id sequences and feature sequences, for micro-scale interest modeling. Each item corresponds to one interest point at the micro-level. MIE also adopts time decay encoding to distinguish the freshness of recent micro-level interests.
	
	\smallskip\noindent\textbf{Graph Clustering with Modularity.} The user-item interaction data can be represented as a bipartite network. The edges~(i.e., interactions) within the bipartite network only exist between user nodes and item nodes. In this paper, we partition the clusters in a user-item bipartite network based on the bipartite modularity~\cite{barber2007modularity}, which is defined as 
	\begin{equation}\label{key}
		Q = \frac{1}{E} \sum_{i,j} \left( A_{ij}  - P_{ij} \right) \delta(c_i,c_j),
	\end{equation}
	where $ E $ is the total number of edges in the bipartite graph, $ A_{ij} $ is the adjacency matrix where the element equals one if an interaction between $ i $ and $ j $ exists, $ P_{ij} $ refers to the expected edge between $ i $ and $ j $ in a graph partitioned by different clusters, and $ \delta(c_i,c_j)$ is the indicator function which equals one if $ i $ and $ j $ belongs to the same cluster, otherwise zero. 
	% The higher Q, the more do the data support the division of a network into modules. 
	%	Bipartite modularity indicates the normalized value of actual edges minus expected edges in clusters. 
	A larger value of $ Q $ means that there are more edges in clusters than expected, which implies a stronger cluster structure.
	The graph modularity $ Q $ can be optimized in an iterative manner according to the Louvain algorithm~\cite{feng2020improving}. After the algorithm converges, the items are grouped into different clusters which are used as the foundation of macro-level interest modeling.
	
	\smallskip\noindent\textbf{Macro-Level User Interest.} For a given user $ u $, we first classify its behavior items into several interest points according to their belonged clusters. Each interest point represents the user's one aspect of macro-level interest. We present an example in Figure~\ref{fig:MPAD}. Given a user $ u $ with full behavior sequence $ I^u = \{ i_1, i_2, i_3, i_4, ..., i_N\} $, we partition these behavior items into four interest points, i.e., $ C^u = \{ c_1, c_2, c_3, c_4\} $. Here $ c^u_1 = \{i_1, i_3\} $ due to that $ i_1$ and $i_3$ belong to the same cluster. 
	We obtain the representation of one interest point by pooling over the embedding of its contained items: 
	\begin{equation}\label{eq:interestPointAgg}
		\mathbf{h}^u_{m} = \text{Aggregate} \left\lbrace \mathbf{e}_{i_x},~~\forall i_x \in c_m \right\rbrace,
	\end{equation}
	where $ c_m $ refers to the $ m $-th interest point, $ \mathbf{e}_{i_x} $ denotes the embedding of behavior item $ i_x $ and $ \text{Aggregate}(\cdot) $ is an aggregation function, which is sum pooling in this paper. 
	
	Then, we perform multi-head attention among the top-M interest groups to obtain the representation of macro-level interests.
	Formally, the formulation of one single-head attention can be written as
	\begin{equation}\label{eq:self_att}
		\text{Att}(\bm{Q}_m,\bm{K}_m,\bm{V}_m) = \text{Softmax}\left(\alpha \bm{Q}_m\bm{K}_m^T \right)\bm{V}_m,
	\end{equation}
	where $ \bm{Q}_m \!=\! \mathbf{h}^u_{m}\bm{W}^Q $, $ \bm{K} \!=\! \mathbf{h}^u_{m}\bm{W}^K $, and $ \bm{V} \!=\! \mathbf{h}^u_{m}\bm{W}^V $ are the linear transformations applied to the representation of interest group $ \mathbf{h}^u_{m} $. 
	% 不同的interest group分别与target item做target attention?
	The scaling factor $ \alpha $ is usually set to be $ 1/\sqrt{d} $ with $ d $ being the dimension of the embedding vector.
	Then, the representation of macro-level user interest via multi-head attention is
	\begin{subequations}\label{eq:dotTALong}
		\begin{align}
			\text{Head}_m &= \text{Att}(\bm{Q}, \bm{K}_m, \bm{V}_m), \\
			\mathbf{h}^u_{\text{macro}} &= \text{Concat}(\text{Head}_1,\cdots,\text{Head}_m)\bm{W}^O,  
		\end{align}
	\end{subequations}
	where $ \text{Concat}(\cdot) $ denotes the concatenation of embedding vectors and $\bm{W^O}$ denotes the linear projection matrix and scales with the number of used heads.
	
	\smallskip\noindent\textbf{Micro-Level User Interest.} 
	User's micro-level interests are usually more dynamic and more concrete than macro-level interests. Therefore, we directly perform multi-head attention towards the individual behavior items, instead of clusters, to obtain the representation of micro-level user interests.
	Noticeably, we inject the time decay corresponding to each behavior item into the embedding to describe the freshness of this aspect of interest. To be more specific, the expanded embedding of each behavior item can be written as
	\begin{equation}\label{key}
		\tilde{\bm{e}}_{i_x} = \text{Concat}\{ \bm{e}_{i_x}, \bm{t}_{i_x} \}, ~~\forall i_x \in I^u_{\text{micro}},  
	\end{equation}
	where $ I^u_{\text{micro}} $ denotes the set of individual behavior items and $ \bm{t}_{i_x} $ is a learnable embedding that represents the time interval from the interaction time till now. 
	Then, we obtain the representation of the user's micro-level interest in the target item as 
	%	performs multi-head target attention between the embeddings of user's short-term behavior item and the target item to model user's short-term interest on the target item, i.e., 
	\begin{subequations}\label{eq:dotTAShort}
		\begin{align}
			\text{Head}_i &= \text{Att}(\bm{Q}_i, \bm{K}_i, \bm{V}_i), \\
			\mathbf{h}_{\text{micro}} &= \text{Concat}(\text{Head}_1,\cdots,\text{Head}_h)\bm{W}^O,  
		\end{align}
	\end{subequations}
	where $ \bm{Q}$, $ \bm{K}$, and $\bm{V} $ follows similar definition as in~\eqref{eq:self_att} but replace the embedding of interest group $ \mathbf{h}^u_{m} $ with the embedding of each individual behavior item $ \tilde{\bm{e}}_{i_x} $. 
	
	\subsection{Context-Aware Accuracy Estimation}
%	We concatenate all user-side and item-side features as input and feed them into multiple MLP layers and a sigmoid function to predict the click probability of a target user $ u $ towards a target item $ i $, which can be formulated as
%	\begin{subequations}
%	\begin{align}
%	&\mathbf{h}^u_{\text{all}} = \text{Concat}\left(\mathbf{e}_i, \mathbf{h}^u_{\text{long}}, \mathbf{h}^u_{\text{short}}, \mathbf{h}^u_{\text{profile}}, \mathbf{h}_{\text{context}} \right), \\
%	&\hat{Y}_{I}(u,i) = \text{Softmax} \left( \text{MLP} \left( \mathbf{h}^u_{\text{all}} \right) \right),
%	\end{align}
%	\end{subequations}
%	where $ \mathbf{h}^u_{\text{short}} $ is user's short-term interest representation, $ \mathbf{h}^u_{\text{long}} $ is user's long-term interest representation, $ \mathbf{e}_i $ is the target item embedding, $ \mathbf{h}^u_{\text{profile}} $ is the user profile embedding, and $ \mathbf{h}_{\text{context}} $ is the context embedding. 
	In this section, we propose the CAE model to refine the point-wise accuracy scores produced by models in the ranking stage into context-aware accuracy scores for the measurement of $ \text{Acc}(u,S) $. 
	The proposed model only performs a linear transformation on the embedding vectors such that it is low-cost for online inference. 
	A brief workflow of CAE is presented in Figure.~\ref{fig:MPAD}.
%	Specifically, the proposed M2 model refines the score of a given item $ i \in S $ based on  Note that the point-wise estimation scores are used as input of M2, such that M2 can be lightweight since it only needs to use the additional features appeared in the re-ranking stage.

	CAE maintains two embeddings to describe the context information. 
	First, when determining the $ k $-th item in a page, we represent the context of previous reviewed items as
	\begin{equation}\label{eq:prevContextAgg}
		\mathbf{h}_{\text{prev}} = \text{Aggregate} \left( \mathbf{e}_i,~~i\in [k-1] \right),
	\end{equation}
	where $ [k-1] = \{1,2,\cdots,k-1\} $.
%	This previous-context embedding represents the previous items in front of the current slot to allocate.
	Second, we represent the context of all candidate items to reflect the overall tendency from ranking models, which can be written as
	\begin{equation}\label{eq:candContextAgg}
	\mathbf{h}_{\text{cand}} = \text{Aggregate} \left( \mathbf{e}_i,~~i \in [N] \right).
	\end{equation}
	
%	The list-wise context is aggregated in the previous-context embedding and the candicate-context embedding. 
	Next, we model the influence from the context of previous items and candidate items towards the target item based on an excitation mechanism proposed in SENet~\cite{hu2018squeeze}. 
%	we employ a similar  which is able to map the list-wise context embedding into a set of channel weights.
	Taking the context of previous items as an example, we obtain its excited representation as
	\begin{equation}\label{key}
		\mathbf{W}_{\text{prev}} = \sigma (\mathbf{W}_2 \cdot \delta (\mathbf{W}_1 \cdot  \mathbf{h}_{\text{prev}}) ),
	\end{equation}
	where $ \sigma(\cdot) $ is the sigmoid activation function, $ \delta(\cdot) $ is the Relu function, $ \mathbf{W}_1 $ and $ \mathbf{W}_2 $ are the linear transformation matrices.
	This excitation operator can be understood as a low-cost attention mechanism to extract the key information embedded in the context vector $ \mathbf{h}_{\text{prev}} $.
	Then, we multiply $ \mathbf{W}_{\text{prev}} $ with the target item embedding to emphasize the influence from the context of previous items:
	\begin{equation}\label{eq:excitation}
		\bm{h}^{i_t}_{\text{prev}} = \mathbf{W}_{\text{prev}} \otimes \bm{e}_{i_t},
	\end{equation}
	where $ \otimes $ denotes the dot product operation. 
	The same goes for the excited vector for the context of candidate items. 
	Moreover, by replacing $ \bm{e}_{i_t} $ in~\eqref{eq:excitation} with the macro-level user interest $ \mathbf{h}^u_{\text{macro}} $ and the micro-level user interest $ \mathbf{h}^u_{\text{micro}} $, we obtain another four excited vectors, i.e., $ \mathbf{h}^u_{\text{pr,lo}} $, $ \mathbf{h}^u_{\text{pr,sh}} $, $ \mathbf{h}^u_{\text{ca,lo}} $, and $ \mathbf{h}^u_{\text{ca,sh}} $, to model the drift of user interest based on the list-wise context. 
%	Then, we perform vector addition, subtraction and dot product target item $ e_i $ and each of the previous-context embedding, candidate-context embedding, long-term interest representation and short-term interest representation. For example, for previous-context embedding, we have
%	\begin{equation}\label{eq:prevContextAgg}
%		\tilde{\mathbf{h}}^u_{\text{before}} = \text{Concat}(\mathbf{h}_{\text{before}}\oplus\mathbf{e}_i || \mathbf{h}_{\text{before}}\otimes\mathbf{e}_i || \mathbf{h}_{\text{before}}\ominus\mathbf{e}_i),
%	\end{equation}
%	where $ \oplus $, $ \otimes$ and $ \ominus $ denotes the addition, subtraction and dot product of embedding vectors, respectively.
	Finally, we concatenate all excited embeddings together and feed it into an MLP layer with softmax function to get the output scores:
	\begin{subequations}
		\begin{align}
		&\mathbf{h}^u_{\text{all}} = \text{Concat} \left(  \mathbf{e}_{i_t}, \bm{h}^{i_t}_{\text{prev}}, \bm{h}^{i_t}_{\text{cand}},  \mathbf{h}^u_{\text{pr,lo}}, \mathbf{h}^u_{\text{pr,sh}}, \mathbf{h}^u_{\text{ca,lo}}, \mathbf{h}^u_{\text{ca,sh}} \right), \\
		&\hat{Y}_{II}(u,i) = \text{Softmax} \left( \text{MLP} \left( \mathbf{h}^u_{\text{all}} \right) \right),
		\end{align}
	\end{subequations}
	This CAE model can be trained with the commonly used cross-entropy loss as in other ranking models.
%	This CAE model can be trained with the cross-entropy loss:
%	\begin{equation}\label{eq:loss}
%			J = \sum\nolimits_{u,i \in \mathcal{D}} \Big( y_{u,i} \log \hat{y}_{u,i} + (1-y_{u,i}) \log (1-\hat{y}_{u,i}) \Big),
%		\end{equation}
%	where $ \mathcal{D} $ denotes the training dataset, $ y_{u,i} $ denotes the real user-item iteration label (e.g., user-item clicks), and $ \hat{y}_{u,i} $ denotes the predicted score by CAE.
	
	\subsection{Perception-Aware Diversity Kernel}
	\label{sec:PDK}
	This section introduces the design of diversity kernel $ \boldsymbol{D}^u_S $ in~\eqref{eq:logProbability}.
	In general, the diversity kernel determines how to evaluate the similarity between any given pairs of items in set $ S $. The elements of $ \boldsymbol{D}^u_S $ determines the $ \log(d^2_i) $ term in~\eqref{eq:finalSelectionStep}.
%	Kernel function design is critical to our proposed method, as it encodes the prior information of the item selection process. 
	Different definitions of the diversity kernel lead to disparate diversification results. 
	In this paper, we introduce the user's multi-scale interests obtained in Sec.~\ref{sec:interest} into the measurement of item similarity. This connects diversity measurement with the user's personal perception of diversity due to distinct interests. 
%	measure the similarity among different items based on the  vectors obtained in Sec.~\ref{sec:interest} to model users' potential diversification requirements on different scales of interests. For example, user may tend to see more items from short-term interest and less items from long-term interest, or otherwise.
	
	Specifically, we define an elementary kernel based on the form of SE kernel~\cite{RW06} for the perception on macro-level interests as
	\begin{equation}\label{eq:longInterestKernel}
		D^u_{\text{macro}}(i,j|E_u) = a^2_{l} \exp \left[ - \frac{\bm{h}^u_{i,\text{macro}}\otimes\bm{h}^u_{j,\text{macro}}}{b^2_{l}} \right],
	\end{equation}
	where $ a^2_{l} $ is the magnitude of the correlated components, $ b_{l} $ is its length scale, and $ \bm{h}^u_{i,\text{macro}} $ refers to the dot product between the item embedding $\bm{e}_{i} $ and the macro-level interest vector $ \bm{h}^u_{\text{macro}} $. Similar goes for $ \bm{h}^u_{j,\text{macro}} $.
%	Note that the learned representations $ \bm{h}^u_{i,\text{long}} $ is related with the user information, such that the constructed kernel is distinct to different users, thereby leading to personalized diversification results.
	The kernel for the perception on micro-level interests can be defined as
	\begin{equation}\label{eq:shortInterestKernel}
		D^u_{\text{micro}}(i,j|E_u) = a^2_{s} \exp \left[ - \frac{\bm{h}^u_{i,\text{micro}}\otimes\bm{h}^u_{j,\text{micro}}}{b^2_{s}} \right],
	\end{equation}
	where $ a^2_{s} $ and $ b_{s} $ are hyper-parameters for micro-level interests.
	We also define another elementary kernel that directly compares the similarity between items based on their embeddings. In this case, the item-level diversity used in existing literature~\cite{lin2022feature,wilhelm2018practical,huang2021sliding,chen2018fast} can be treated as a special case of PDK. In particular, this kernel can be defined as
	\begin{equation}\label{eq:itemEmbKernel}
		D_{\text{item}}(i,j|E_u) = a^2_{s} \exp \left[ - \frac{\bm{e}^u_{i}\otimes \bm{e}^u_{j}}{b^2_{s}} \right],
	\end{equation}
	
	These elementary kernels can be merged into a composite kernel without influencing the kernel properties via addition and multiplication operations~\cite{RW06}. More complicated operations such as automatic kernel learning are also worth trying for better adaptivity and full automation of a system, e.g., deep DPP~\cite{gartrell2018deep}, which can be explored in the future. 
	In this work, we adopt the addition operation to construct this composite kernel:
	\begin{equation}\label{key}
		D^u(i,\!j|E_u) \!=\! D^u_{\text{item}}(i,\!j|E_u) + \beta_1 \!\cdot\! D^u_{\text{macro}}(i,\!j|E_u)  + \beta_2 \!\cdot\! D^u_{\text{micro}}(i,\!j|E_u),
	\end{equation}
	where $ \beta_1 $ and $ \beta_2 $ are the hyper-parameters to control the influence from macro-level and micro-level interest to the diversification results. 
	
 	\begin{figure}
		\centering
		\includegraphics[trim = 5 5 5 5, clip, width=1\columnwidth]{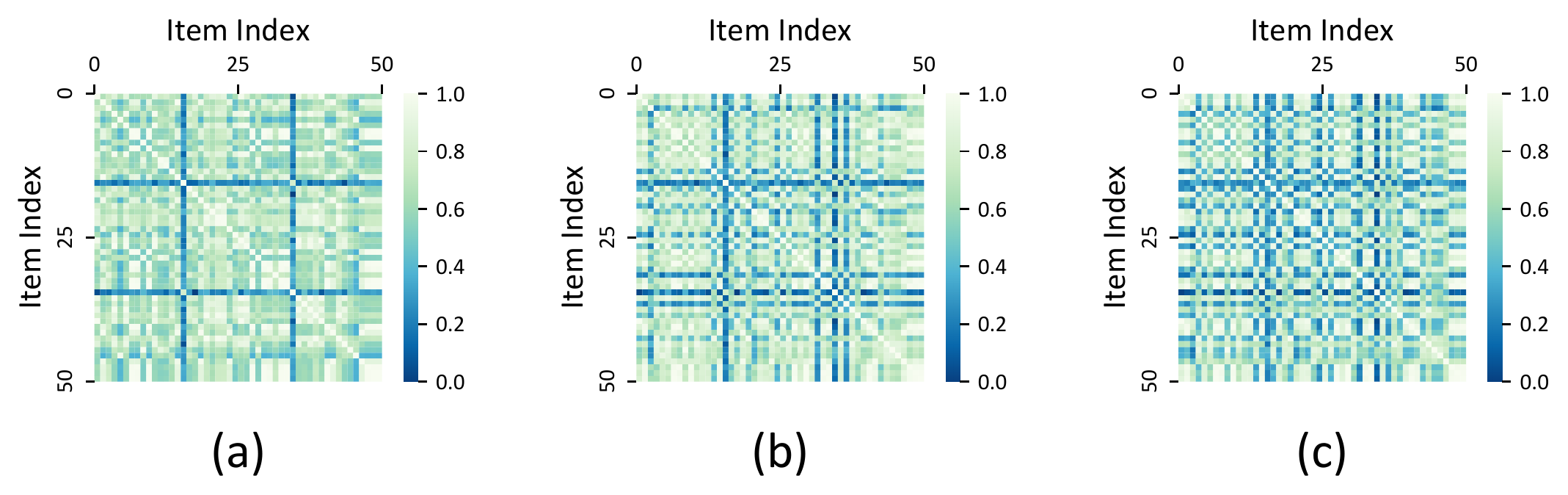}
		\caption{Influence of adding multi-scale user interests to the diversity measurement. (a) item-level similarity; (b) item-level similarity with macro-level interests; (c) item-level similarity with macro-level and micro-level interests.}
		\label{fig:PDK}
	\end{figure}
	We give an example in Figure.~\ref{fig:PDK} to show the change of diversity measurement when adding user interests into the kernel. Figure.~\ref{fig:PDK}(a) shows the item similarity in $ D^u_{\text{item}}(i,\!j|E_u) $ which only considering the distance of item embedding. Figure.~\ref{fig:PDK}(b) shows the item similarity after adding the macro-level interests $ D^u_{\text{macro}}(i,\!j|E_u) $ into the kernel. Figure.~\ref{fig:PDK}(c) shows the item similarity of the complete kernel $ D^u(i,\!j|E_u) $. It is clear that part of dissimilar items transforms into similar items due to the consideration of user interests, and vice versa. In this way, the similarity values of the same set of items are different for distinct users, thereby leading to perception-aware diversification.
%	Other ways to construct the composite kernel can be explored in the future.
	
	\subsection{Online Implementation}
	\label{sec:implementation}
	\begin{figure}[tb]
		\centering
		\includegraphics[trim = 18 18 18 15, clip, width=0.9\columnwidth]{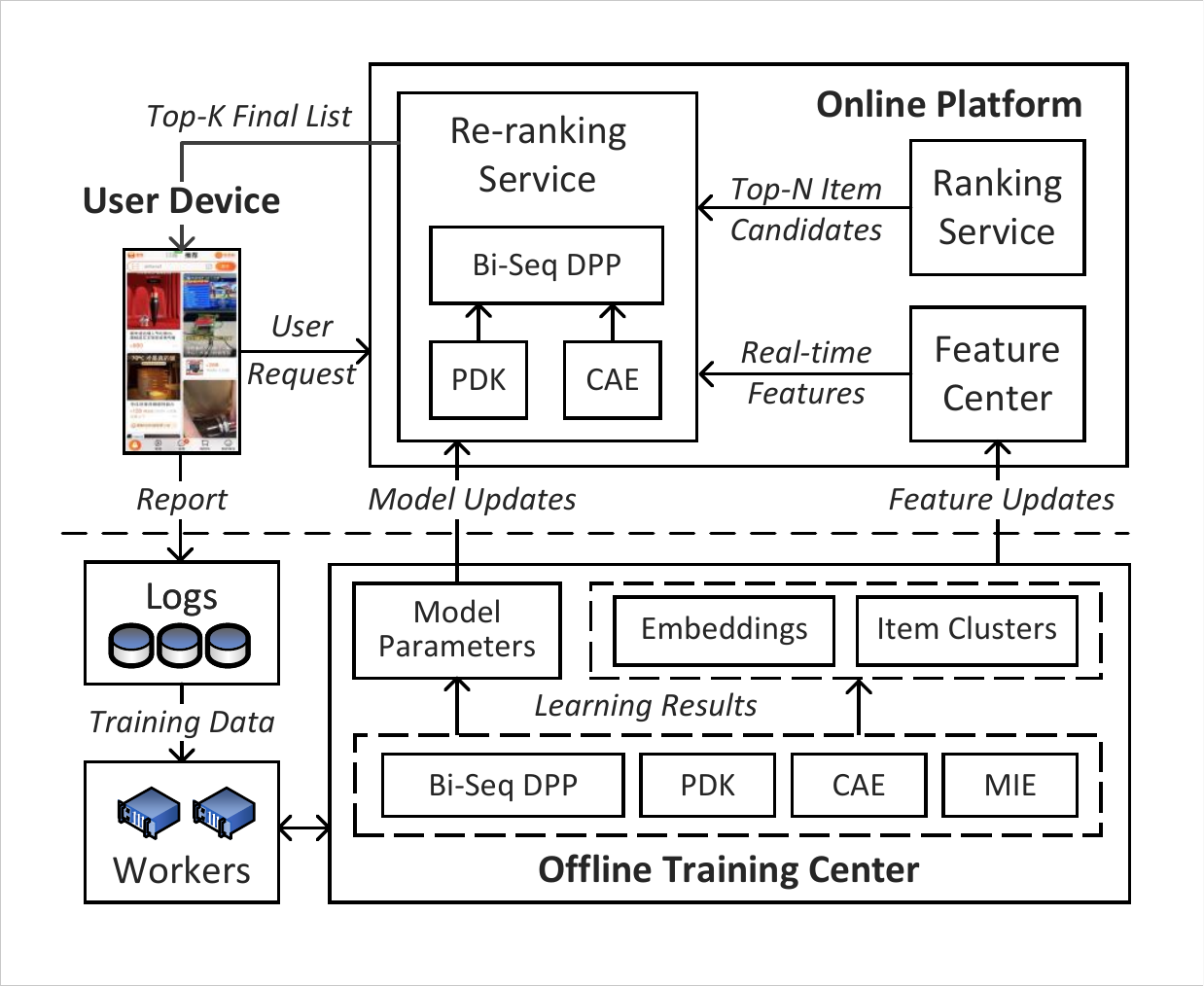}
		\caption{The System Architecture for Online Deployment}
		\label{fig:onlineArch}
	\end{figure}
	In this section, we introduce the online implementation of MPAD in the Homepage Feed of Taobao Mobile App.
	The presented system architecture is able to handle $ 120,000 $ QPS at traffic peak and respond within $ 20 $ milliseconds in general. It now serves the main traffic of Taobao to provide services to hundreds of millions of users towards billions of items in Taobao every day.
	
	The architecture to implement the proposed MPAD model in Taobao is presented in Fig.~\ref{fig:onlineArch}, including the workflow of both offline training and online serving.
	The offline training is based on a distributed machine learning platform. The learned embedding and item clustering results are uploaded to the feature center for online serving. The re-ranking service retrieves user and item features from the feature center in real-time and feeds them into a series of models to determine the final item list. Note that the graph clustering in MIE only performs offline to reduce online complexity.
	
	The online inference complexity consists of three parts.
	First, each element in PDK requires the computation of dot product between two embedding vectors which incurs a complexity of $ \mathcal{O}(d) $ where $ d $ is the length of embedding, such that the overall complexity of computing PDK scales as $ \mathcal{O}(dN^2) $, where $ N $ is the number of candidate items.
	Note that the dot product of embedding vectors between hot items and active users can be pre-computed offline to save a lot of computations.
	Second, the accuracy estimation in CAE only involves linear transformation over embedding vectors in the exciting mechanism and the final MLP layer. As such, the complexity scales linearly with the length of embedding vectors, i.e., $ \mathcal{O}(dN) $ where we assume the length of each input embedding is $ d $.
	Third, the BS-DPP runs in the same complexity as standard DPP~\cite{chen2018fast}, i.e., $O(K^3)$ time for unconstrained MAP inference and $O(K^2N)$ to return $ K $ items.
%	the sequential item selection process in BS-DPP.
	% complexity of BS-DPP
%	In the $i$-th iteration of BS-DPP, for each item $i\in I\setminus S$, updating ${\bf c}_i$ and $d_i$ involve the inner product of two vectors of length $d$, resulting in overall complexity $O(dK)$ where $ K $ is the number of items to return.
%	Therefore, the BS-DPP runs in $O(K^3)$ time for unconstrained MAP inference and $O(K^2N)$.

	\section{Experimental Results}
	In this section, we conduct extensive experiments on both offline datasets and real-world online RS with the goal to answer the following research questions.
	\smallskip\noindent\textbf{Q1:} Does MPAD outperform other SOTA methods in terms of accuracy and diversity for feed recommendation? 
	\smallskip\noindent\textbf{Q2:} How do different components of MPAD influence the final performance?
	\smallskip\noindent\textbf{Q3:} How does MPAD perform in real-world feed recommendation platforms?
	\subsection{Experimental Setup}
	\subsubsection{Datasets}
	We conduct offline experiments on three public available datasets: MovieLens dataset\footnote{\url{https://grouplens.org/datasets/movielens/10m/}}, Wechat dateset\footnote{\url{https://algo.weixin.qq.com/2021/intro}}, and Taobao dataset\footnote{\url{https://tianchi.aliyun.com/dataset/dataDetail?dataId=56}}. 
	Specifically, MovieLens dataset is a widely-used benchmark dataset for movie recommendations, which contains $ 10 $ million samples. Here we propose it for easy reproduction.
	Wechat dataset is collected from $ 7.3 $ million of video playback logs on Wechat Mobile App. It is one of the largest mobile social applications in China. The dataset involves $ 20,000 $ users and $ 96,564 $ videos. The label is marked as positive if the user has watched more than 90\% playback progress of a video.
	Taobao dataset is a widely used public benchmark dataset for online advertising, which contains over $ 100 $ million ad display/click logs collected from Taobao Mobile App. It is one of the largest online merchandise applications in China. The logs involve $ 1 $ million users and $ 800 $ thousand of ads collected on Taobao Mobile App. 
%	Users’ recent $ 200 $ behaviors are also provided in the logs.

	\subsubsection{Comparing Methods}
	We compare MPAD with both point-wise and list-wise mainstream methods for recommendation tasks. 
%	Point-wise methods predict the interaction probability, e.g., click or pay, for any given user-item pair based on the raw features derived from user and item profile. 
%	list-wise methods predict the interaction probability of all items in a page by making use of both the context information and the raw features. 
	
	\textbf{Point-wise baselines}: we compare with four commonly used point-wise baselines, i.e., the shallow model based on linear regression~(LR)~\cite{mcmahan2013ad}, the PNN model~\cite{qu2016product} which performs feature interaction with different product operations, the Wide \& Deep learning model~(WDL)~\cite{cheng2016wide} and DeepFM model~\cite{guo2017deepfm} which adopt a hierarchical structure consists of linear and deep layers. 
	We also compare with a few representative user interest models, i.e., DIN~\cite{zhou2018deep} which models short user behavior sequences with the target attention mechanism; DIEN~\cite{zhou2019deep} which uses
	an interest extraction layer based on Gated Recurrent Unit~(GRU) to model users’ temporal drifting  interest; SIM~\cite{qi2020search} which models user's full behavior sequence based on a two-stage paradigm.
	
	\textbf{List-wise baselines}: we compare with three representative list-wise baselines, i.e., DLCM~\cite{ai2018learning} which applies GRU to encode the input ranking list, accompanied with a global vector to learn a powerful scoring function for list-wise re-ranking; PRM~\cite{pei2019personalized} which uses the self-attention mechanism to capture the mutual influence among items in the input ranking list; Seq2Slate~\cite{bello2018seq2slate} which adopts RNN and pointer network to encode the previous selected items when selecting the most appropriate item for next step. We compare with the statistical models, i.e., maximal marginal relevance (MMR)~\cite{carbonell1998use} and fast DPP~\cite{chen2018fast}. Both of them have a tunable parameter to balance accuracy and diversity, similar to MPAD. 
	We also compare with the generative-based models which directly generate item lists as the final results, including ListCVAE~\cite{jiangCVAE} and PivotCVAE\cite{liu2021pivotcvae}.

	\subsubsection{Metrics.}
	For accuracy estimation, we use the commonly used Area Under ROC (AUC) and Logloss (cross entropy) to evaluate the point-wise estimation performance; and use normalized discounted cumulative gain (nDCG)~\cite{jarvelin2017ir} and mean average precision~(MAP) to measure the list-wise estimation performance. nDCG@K or MAP@K refers to the performance of top-k recommended items in the return list.
%	AUC (Area Under the ROC curve) and Logloss (cross entropy). These two metrics evaluate the performance from two different angels: AUC measures the probability that a positive instance will be ranked higher than a randomly chosen negative one. It only takes into account the order of predicted instances and is insensitive to class imbalance problem. Logloss, in contrast, measures the distance between the predicted score and the true label for each instance.
	For list-wise diversity, we use intra-list average distance (ILAD)~\cite{zhang2008avoiding} to evaluate the richness of diversified items in a page.
%	, which is defined as $ \text{ILAD} = \mathop{\text{mean}}\nolimits_{u \in U} \mathop{\text{mean}}_{i,j \in R_u, i\neq j} (1- S_{ij}) $.
	Moreover, we use PV, Stay Time, Category Breadth, CLICK, CTR, and GMV to evaluate online performance. Here, PV refers to the total number of browsed items, Stay Time is the average browsing time of all users, and Category Breadth computes the average number of distinct categories of all exposed items on all pages, reflecting the diversity of recommendation results.
	CLICK refers to the total number of clicked items, CTR equals CLICK/PV which measures users’ willingness to click. GMV is a term used in online retailing to indicate the total sales monetary value for merchandise sold over a certain period of time. We use the time period of a complete day for all online metrics in this paper.
	
	\subsubsection{Parameter Settings.}
	\label{sec:parameter_setting}
	In all experiments, we use the validation set to tune the hyper-parameters to generate the best performance for different methods. The learning rate is searched from $10^{-4}$ to $10^{-2}$. The L2 regularization term is searched from $10^{-4}$ to $1$. All models use Adam as the optimizer. We extract micro-level interests from the user's recent 100, 50, and 20 behavior items for MovieLens, WeChat, and Taobao, respectively. For macro-level interests, we group all items into 20, 241, and 6769 clusters for MovieLens, WeChat, and Taobao, respectively. We assign each user's recent behavior to these clusters, and we select the top-5 interest groups to compute their macro-level interests.
	
	\subsection{Offline Evaluation}
	
	\begin{table}[]
		\centering
  		\caption{Comparison of user interest modeling~(bold: best; underline: runner-up). The marker * denotes that our model performs significantly better than the runner-up with $p<0.01$ over 25 runs. }
		\resizebox{0.95\columnwidth}{!}{
			\begin{tabular}{@{}ccccccc@{}}
				\toprule
				\multirow{2}{*}{Method} & \multicolumn{2}{c}{MovieLens}     & \multicolumn{2}{c}{WeChat}        & \multicolumn{2}{c}{Taobao}        \\ \cmidrule(l){2-7} 
				& AUC ($\uparrow$)         & Logloss ($\downarrow$)      & AUC  ($\uparrow$ )        & Logloss ($\downarrow$)       & AUC   ($\uparrow$ )       & Logloss  ($\downarrow$)      \\ \midrule
				LR     & 0.7337       & 0.6254       & 0.6433       & 0.6656       & 0.5725       & 0.1930       \\
				PNN    & 0.7836       & 0.5597       & 0.6976       & 0.6295       & 0.6309       & 0.1896       \\
				WDL    & 0.7883       & 0.5584       & 0.6968       & 0.6295       & 0.6316       & 0.1894       \\
				DeepFM & 0.7894       & 0.5571       & 0.6979       & 0.6290       & 0.6315       & 0.1898       \\
				DIN    & 0.8024       & 0.5394       & 0.6951       & 0.6319       & {\ul 0.6333} & 0.1896       \\
				DIEN   & {\ul 0.8028} & {\ul 0.5344} & 0.6994       & {\ul 0.6290} & 0.6324       & 0.1902       \\
				SIM    & 0.8023       & 0.5349       & {\ul 0.7006} & 0.6309       & 0.6312       & {\ul 0.1893} \\ \midrule
				Ours   & $ \textbf{0.8056}^* $ & $ \textbf{0.5305}^* $ & $ \textbf{0.7014}^* $ & $ \textbf{0.6279}^* $ & $ \textbf{0.6361}^* $ & $ \textbf{0.1884}^* $ \\ \bottomrule
			\end{tabular}
		}
		\label{tab:multiScaleInterst}
	\end{table}

	\begin{table}[]
		\centering
		\caption{Comparison of item quality in the item list~(bold: best; underline: runner-up).  The marker * denotes that our model performs significantly better than the runner-up with $p<0.01$ over 25 runs. }
		\resizebox{.92\columnwidth}{!}{
		\begin{tabular}{@{}cccccc@{}}
			\toprule
			Dataset                    & Model     & NDCG@3          & NDCG@10         & MAP@3           & MAP@10          \\ \midrule
			\multirow{9}{*}{MovieLens} & DIN       & 0.9017          & 0.9230          & 0.8760          & 0.8893          \\
			& DIEN      & 0.9031    & 0.9239 & 0.8775  & 0.8900  \\
			& SIM       & 0.9004  & 0.9219  & 0.8747  & 0.8879  \\ 
			\cmidrule(l){2-6}
			& Seq2Slate & 0.9098  & 0.9295  & 0.8863  & 0.8978  \\
			& DLCM      & 0.9095  & 0.9293  & 0.8857  & 0.8976  \\
			& PRM       & {\ul 0.9102}  & {\ul 0.9296}  & {\ul 0.8865}  & {\ul 0.8981}  \\ 
			\cmidrule(l){2-6}
			& ListCVAE  & 0.8349  & 0.8800  & 0.8154  & 0.8632 \\
			& PivotCVAE & 0.8608  & 0.8928  & 0.8423  & 0.8775  \\ 
			\cmidrule(l){2-6}
			& Ours      & $\textbf{0.9148}^*$ & $\textbf{0.9332}^*$ & $\textbf{0.8918}^*$ & $\textbf{0.9027}^*$ \\ \midrule
			\multirow{9}{*}{WeChat}    & DIN       & 0.6811  & 0.7200  & 0.6084  & 0.5935  \\
			& DIEN      & 0.6970  & 0.7310  & 0.6256  & 0.6075  \\
			& SIM       & 0.6976  & 0.7329  & 0.6272  & 0.6100  \\ 
			\cmidrule(l){2-6}
			& Seq2Slate & 0.7001  & 0.7342  & 0.6309  & 0.6148  \\
			& DLCM      & {\ul 0.7029}    & {\ul 0.7373}    & {\ul 0.6318}    & {\ul 0.6164}    \\
			& PRM       & 0.7001          & 0.7363          & 0.6280          & 0.6152          \\ 
			\cmidrule(l){2-6}
			& ListCVAE  & 0.5710  & 0.6533  & 0.4975  & 0.5333  \\
			& PivotCVAE & 0.5738  & 0.6547  & 0.4993  & 0.5362  \\
			\cmidrule(l){2-6}
			& Ours      & $\textbf{0.7095}^*$ & $\textbf{0.7419}^*$ & $\textbf{0.6398}^*$ & $\textbf{0.6216}^*$ \\ \midrule
			\multirow{9}{*}{Taobao}    & DIN       & 0.2017  & 0.3172  & 0.1654  & 0.2227  \\
			& DIEN      & 0.2003  & 0.3155  & 0.1631  & 0.2202  \\
			& SIM       & 0.2006  & 0.3200  & 0.1645  & 0.2237  \\ 
			\cmidrule(l){2-6}
			& Seq2Slate & 0.2093          & 0.3294          & 0.1728          & 0.2326          \\
			& DLCM      & 0.2115          & {\ul 0.3308}    & 0.1749          & {\ul 0.2337}    \\
			& PRM       & {\ul 0.2118}    & 0.3303          & {\ul 0.1750}    & 0.2335          \\ 
			\cmidrule(l){2-6}
			& ListCVAE  & 0.1767  & 0.3076  & 0.1523  & 0.2193  \\
			& PivotCVAE & 0.1785  & 0.3120  & 0.1512  & 0.2224  \\
			\cmidrule(l){2-6}
			& Ours      & $\textbf{0.2166}^*$ & $\textbf{0.3339}^*$ & $\textbf{0.1799}^*$ & $\textbf{0.2372}^*$ \\ \bottomrule
		\end{tabular}
	}
	\label{tab:itemQuality}
	\end{table}

	\begin{figure}[t]
		\centering
		\subfigure[Impact of parameter $ \alpha $ on Taobao dataset.]
		{ 	
			\label{fig:diversityA}
			\includegraphics[trim = 8 8 8 8, clip, width=0.95\columnwidth]{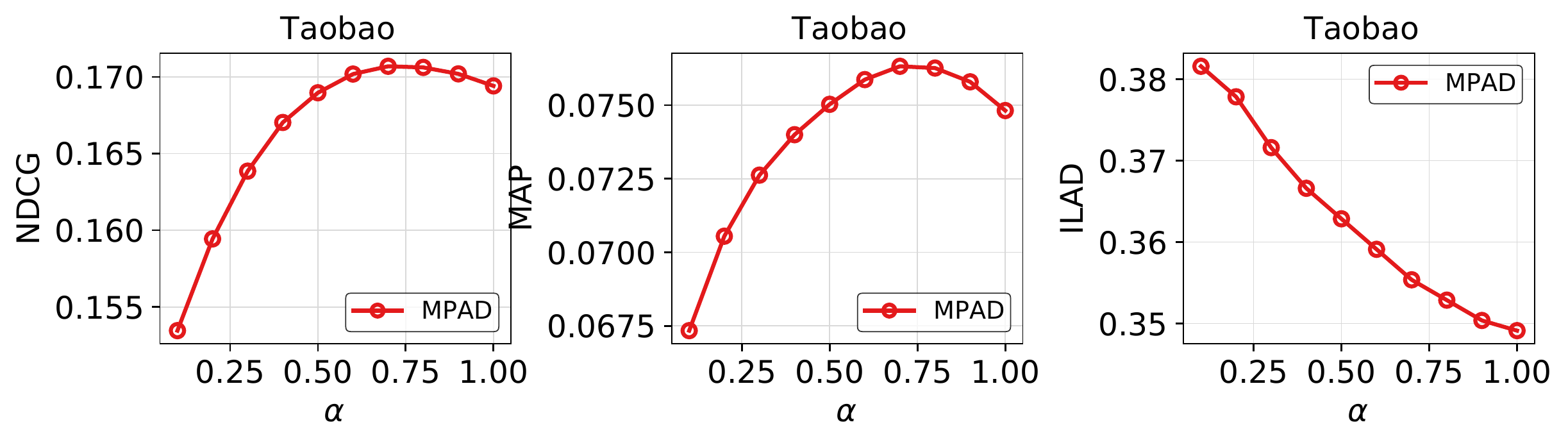}
		}
		\subfigure[Comparison of trade-off performances.]
		{ 	
			\label{fig:diversityB}
			\includegraphics[trim = 8 8 8 8, clip, width=0.95\columnwidth]{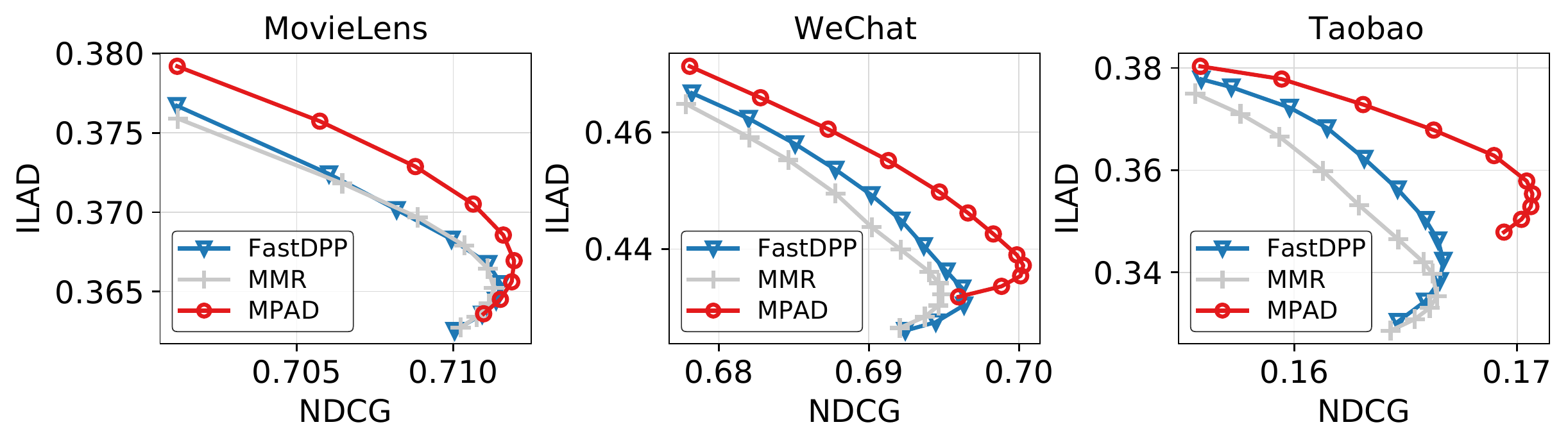}
		}
		\caption{Trade-off between accuracy and diversity.}
		\label{fig:diversity}
	\end{figure}

%    \begin{table}[htbp]
%    \centering
%    \caption{Inference time comparison of rerank models.}
%        \begin{tabular}{c|c|c|c}
%        \toprule
%        Model & Movielens & WeChat & Taobao \\
%        \midrule
%        Seq2Slate & 0.0745  & 0.0801  & 0.1357  \\
%        DLCM  & 0.0584  & 0.0595  & 0.0852  \\
%        PRM   & 0.0457  & 0.0455  & 0.0755  \\
%        ours  & \textbf{0.0436} & \textbf{0.0410} & \textbf{0.0742} \\
%        \bottomrule
%        \end{tabular}%
%    \label{tab:infer_time}%
%    \end{table}%

	This section compares the experimental results of MPAD and other baselines on offline datasets to answer Q1 and Q2. 
	% effectiveness of multi-scale user interest: MIE
	First, we verify the effectiveness of interest modeling in MIE.
	For all datasets, we train MPAD and other competing methods using the same user behavior sequences. 
	The interest modeling techniques differ from each other among the comparing methods.
	In particular, MPAD makes use of the cluster-based interest model proposed in Sec.~\ref{sec:interest}. 
	LR, DeepFM, and WDL treat user behavior sequences as raw features and directly feed them into linear/MLP layers for feature crossing.
	DIN and DIEN adopt TA/GRU units to model short-term user interests. SIM introduces an additional retrieval layer to select top-$ k $ items from user's full behavior sequences to model the lifelong user interest.
%	We also add an ablation study by replacing the cluster-based interest model in MPAD with a DIN-like self-attention layer to examine the benefits from graph clustering.  
	As shown in Table~\ref{tab:multiScaleInterst}, the results verify that MIE outperforms other user interest models remarkably, in terms of both AUC and Logloss. This indicates that MIE is more robust to the disturbing noise hidden in the raw item-level features within behavior sequences that may undermine learning performance. 
%	In this context, our proposed MIE model can alleviate this issue by learning from a more robust cluster-based representation.
	
	% Performance of item accuracy: MIE+CAE
	Next, we compare the performance of accuracy estimation among all point-wise and list-wise ranking methods. For MPAD, we only activate MIE and CAE components for this experiment. 
	As shown in Table~\ref{tab:itemQuality}, the point-wise baselines achieve generally worse performance than the list-wise baselines on all datasets. This verifies that the mutual influence among the input ranking list incurs a great impact on list-wise recommendation. Therefore, it is of vital importance to consider the influence of browsing context in feed recommendations.
	Moreover, our proposed MPAD consistently yields the best performance on all datasets in terms of both NDCG and MAP. This verifies that MPAD has a superior capability to model the contextual influence among consecutive items, due to the modeling of browsing context and the user's multi-scale interests.
	
	% balance of item accuracy & diversity: MIE+CAE+PDK+M0
	Now we examine the capability to balance item accuracy and diversity in MPAD. We activate all components in MPAD for this experiment.
	As shown in Figure~\ref{fig:diversity}, when decreasing the parameter $ \alpha $ in~\eqref{eq:finalSelectionStep}, ILAD decreases monotonously while nDCG increases at first and then decrease a bit. When $ \alpha=0 $, MPAD directly returns items with the highest accuracy scores, regardless of the item diversity. The results indicate that it is critical to introduce a proper amount of diversity into the item list to improve the joint utility of accuracy and diversity for feed recommendation.
	Then, we compare MPAD with MMR and fastDPP. The tunable parameters of all methods are chosen such that different algorithms have approximately the same range of nDCG. 
	The result in Figure~\ref{fig:diversity} shows that, among all comparing methods, our proposed MPAD exhibits the best item accuracy-diversity trade-off performance. This is probably due to the superior performance of accuracy and diversity estimation from MIE, CAE, and PDK. 
	It is also noteworthy that each curve in Figure~\ref{fig:diversity} has an inflection point, corresponding to the optimal balance of accuracy and diversity. In practical applications, the parameter $ \alpha $ should be tuned to reach such an optimal status to deliver the best experience for customers.
	
	\subsection{Online Evaluation}
	\begin{table}[]
		\centering
		\caption{Results of online A/B tests in TaoBao App. }
		\scalebox{0.8}{
		\begin{tabular}{@{}ccccccc@{}}
			\toprule
			& PV     & Breadth & Stay Time & CLICK    & CTR    & GMV    \\ \midrule
			vs Heuristic    & +1.29\% & +4.02\%    & +1.95\%    & +2.38\% & +1.07\% & +1.29\% \\
			vs fastDPP & +0.10\% & +1.46\%    & +1.41\%    & +1.77\% & +1.67\% & +0.27\% \\ \bottomrule
		\end{tabular}
		}
		\label{tab:online}
	\end{table}
%	This section presents the online performance of MPAD to answer Q1.
	MPAD has been fully deployed in the homepage feed of Taobao named \textit{Guess-you-like} to serve the main traffic. 
	In general, Guess-you-like one of the largest merchandise feed recommendation platform in China, which serves more than hundreds of millions of users towards billions of items every day. 
	In Guess-you-like, users can slide to browse and interact with endless items in a sequential manner, as shown in Figure~\ref{fig:feedExample}.
% Once finished reviewing the previous page, the platform will return a fixed-length item list as the new page to users.
	We deploy MPAD at the re-ranking stage in Guess-you-like platform, which takes hundreds of candidate items from the ranking stage as input and outputs a fixed-size item list to form a new page. 
%	Compared with offline experiments, there are relatively richer features for online implementation since we own the complete records on the platform. For example, main features for user-side include: preferred categories, favorite brands, user profile (e.g., age, occupation, and gender), member information, user behavior sequences~(e.g., long-term sequence and short-term sequence), etc; main features for item-side include: category, properties, brand, shop, statistics (e.g., price, trade, and CTR), etc. For all feature processing, model training, and online serving, we follow the architecture presented in Figure~? for implementation. 
	The online performance is compared against the fast DPP method and a heuristic method. 
	Specifically, the fast DPP method uses point-wise ranking scores and item embedding vectors from ranking models as input, similar to~\cite {chen2018fast}.  
	The heuristic method adjusts the item order according to a series of heuristic rules predefined with expert knowledge, e.g., no more than two items within the same category on one screen. It is a commonly used diversification strategy in industrial applications.
	
	The performance in Table~\ref{tab:online} is averaged over two consecutive weeks. We have the following observations.
	Compared with the heuristic method, first, MPAD achieves a performance improvement of $ 2.38\% $ for CLICK, $ 0.62\% $ for CTR, and $ 0.48\% $ for GMV, indicating that our framework is able to increase the user's willingness to interact with the items. The less improvement on GMV is due to that we mainly optimize MPAD towards the CLICK goal to be consistent with the business orientation. It is noteworthy that $ 1\% $ improvement is a considerable enhancement in real-world RS, especially for applications with billion-scale users and items. In Guess-you-like, $ 1\% $ improvement on CLICK brings millions of clicks every day.
	Second, the Category Breadth per page increases by around $ 4\% $ at the same time, which verifies that MPAD is able to promote diversity in the recommended items as well as accuracy.
	Third, the Stay Time increases by $ 1.95\% $ and the PV increases by $ 1.29\% $, which indicates that MPAD can attract users to stay at the platform.  
%	It is worth mentioning that the implementation of MPAD at the re-ranking stage only increases an average cost of $ 15 $ milliseconds for online inference, which would not be a bottleneck for online service. 
    MPAD also outperforms fastDPP in all the above metrics.
	All these improvements verify that MPAD is able to enhance both the item accuracy and diversity in the recommendation results and well balance their trade-off to attract users in feed recommendation.
	
	\subsubsection{Case Study}
 	\begin{figure}
		\centering
		\includegraphics[trim = 5 5 5 5, clip, width=0.95\columnwidth]{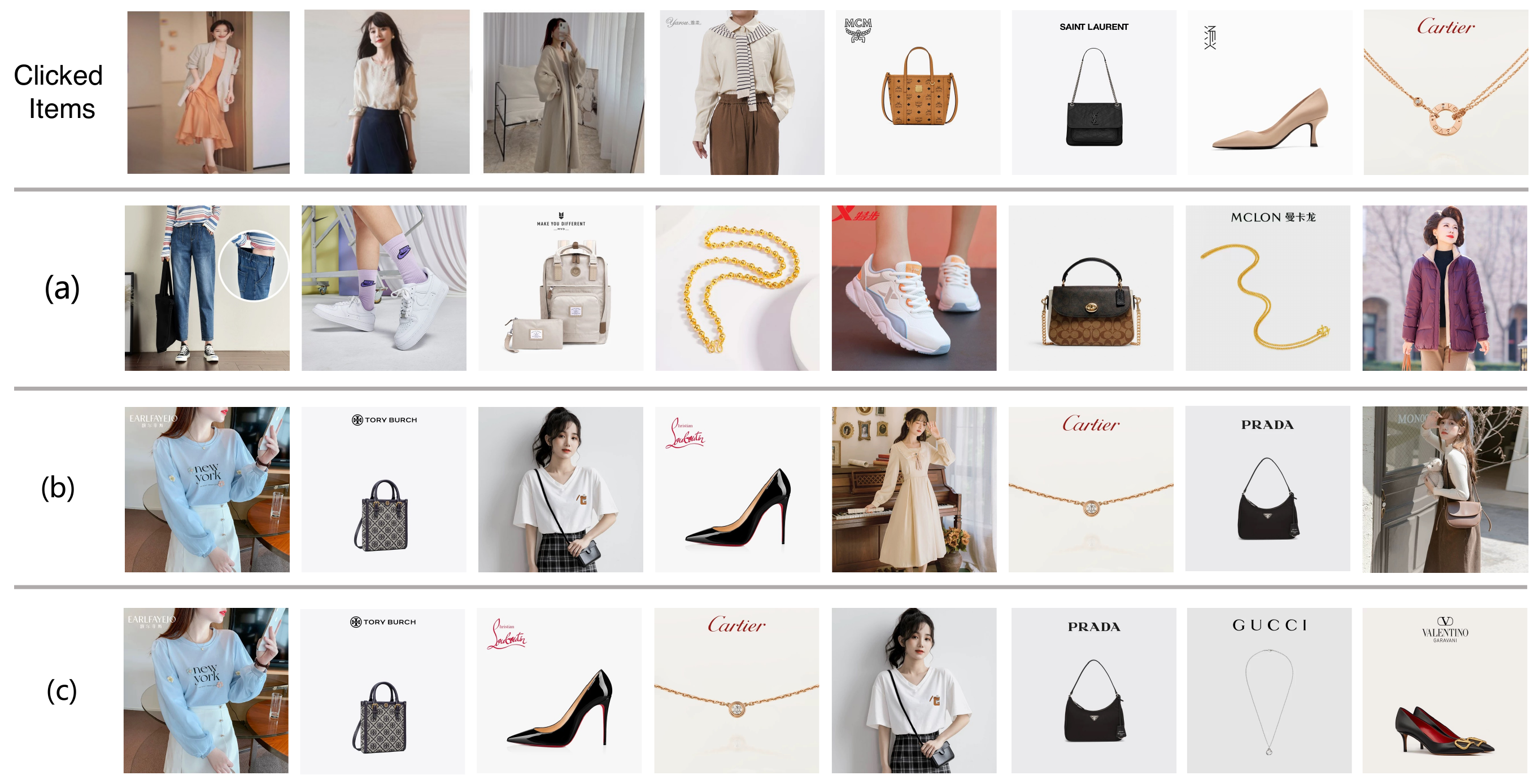}
		\caption{An example of personalized diversification.}
		\label{fig:caseStudy}
	\end{figure}
	In Figure.~\ref{fig:caseStudy}, we present one case to illustrate how MPAD diversifies items to suit personal interests.
	We sample a female customer who recently clicked a series of clothing and dressing items, which indicate her browsing interests. 
	Figure.~\ref{fig:caseStudy}(a) presents the diversified results based on heuristic rules which are universal for all users. It is clear that a few less relevant items appear in the recommendation result, such as sports shoes and down cloth. 
	Figure.~\ref{fig:caseStudy}(b) shows the results obtained by MPAD, where the recommended items are all relevant to the clicked items and are well-spaced to avoid presenting similar items in a row. 
	Figure.~\ref{fig:caseStudy}(c) shows the results of adjusting the parameter $ \alpha $ in MPAD to increase diversity. The items are now more proportioned than those in~(b). For example, the number of clothing items decreases from four in~(b) to only two in~(c), and their distance is greater. 
	This example qualitatively illustrates the effectiveness of MPAD in delivering perception-aware diversification services based on user interests.

	\section{Related Work}
	\smallskip\noindent\textbf{Re-ranking Methods.}
	Traditional point-wise ranking models focus on predicting the interaction label between any given user-item pairs, e.g., Wide\&Deep~\cite{cheng2016wide}, DIN~\cite{zhou2018deep} and SIM~\cite{pi2019practice}, regardless of the context information in a full recommendation list. However, in feed products, the mutual influence between items exhibits a great influence on user behaviors since users are reviewing items in a sequential manner. 
	Recent works on re-ranking propose to consider the mutual influence between items in a list-wise manner, which includes three main research lines, i.e., RNN-based methods, attention-based methods, and evaluator-generator-based methods.  
	% RNN-based re-ranking
	Specifically, the RNN-based methods model the mutual influence based on RNN structures. 
	DLCM~\cite{ai2018learning} uses gated recurrent units (GRU) to sequentially encode the top-ranked items with their feature vectors. MiDNN~\cite{zhuang2018globally} applies the long-short term memory (LSTM), with a global feature extension method to capture cross-item influences. Seq2Slate~\cite{bello2018seq2slate} extends MiDNN by adopting a more flexible pointer network to solve the re-ranking problem.
	% % attention-based re-ranking
	% The attention-based methods, on the other hand, use self-attention mechanism to directly model the interactions between any pair of candidate items, without a sequential encoding structure as in RNN. 
	% PRM~\cite{pei2019personalized} uses a pretrained embedding to extract personalized mutual influences between candidate items and uses multiple blocks of self-attention layers with position encoding to generate list-wise prediction results. 
	% PFRN~\cite{huang2020personalized} proposes a Listwise Feature Encoding (LFE) structure to extract context-aware information from all inputs and model the mutual influences between items, which is based on multi-head self-attention with relative position representations.
	% % generator-evaluator re-ranking
	% The evaluator-generator-based methods propose using a generator to generate feasible permutations and using an evaluator to evaluate their list-wise utility and determine the optimal permutation. Representatives include SEG~\cite{wang2019sequential} and GRN~\cite{feng2021grn}.

    % short version 
    Attention-based methods use self-attention to model item interactions without RNN's sequential structure. PRM~\cite{pei2019personalized} uses pretrained embedding to extract item interactions and generate list-wise predictions with self-attention blocks and position encoding. PFRN~\cite{huang2020personalized} uses Listwise Feature Encoding for context-aware item interaction modeling with multi-head self-attention and relative position representation. Evaluator-generator methods use a generator to generate permutations and an evaluator to determine optimal permutation, e.g., SEG~\cite{wang2019sequential} and GRN~\cite{feng2021grn}. 
	These re-ranking models mainly focus on improving recommendation accuracy instead of a joint utility of both accuracy and diversity.
	
	\smallskip\noindent\textbf{Diversity Methods.}
	It has been widely acknowledged in diversified recommendation methods that accuracy should not be the only goal of recommendation tasks since it may lead to a return of highly similar items to harm user’s satisfaction with the recommendation results~\cite{lin2022feature,huang2021sliding,abdool2020managing,zheng2021dgcn,wilhelm2018practical,chen2018fast,ashkan2015optimal,boim2011diversification,borodin2012max,carbonell1998use,qin2013promoting,sha2016framework}. 
	Research on diversification includes three main streams. 
	% rule-based methods
	The first stream of methods adopts heuristics rules to deal with item order in a post-processing manner. 
	The representative work is maximal marginal relevance (MMR)~\cite{carbonell1998use}, which represents relevance and diversity with independent metrics and maximizes the marginal relevance with a trade-off parameter. 
	Other greedy heuristics methods vary in the definition of this marginal relevance~\cite{ashkan2015optimal,boim2011diversification,borodin2012max,carbonell1998use,qin2013promoting,sha2016framework}. 
	% learning-based methods
	The second stream of methods treats diversified recommendation as an end-to-end learning task. DCF~\cite{cheng2017learning} proposes to solve the coupled parameterized matrix factorization and structural learning problems based on collaborative filtering. 
	BGCF~\cite{sun2020framework} applies bayesian graph convolutional neural networks to model the uncertainty between user-item and bring diversity into recommendation indirectly. 
	DSSA~\cite{jiang2017learning} adopts the attention mechanism to determine the importance of the under-covered subtopics, where the relevance and the diversity are jointly estimated with subtopic attention. 
	% DPP-based methods
	The third stream of methods is based on statistical models. 
	The representative is the determinantal point process~(DPP) which measures set diversity by describing the probability for all subsets of the item set. The maximum a posteriori~(MAP) in DPP to generate diverse lists is NP-hard, such that many related works focus on the approximation of DPP for low-complex iterates.
	For example, Fast DPP~\cite{chen2018fast} proposes a greedy approximation to accelerate the MAP inference for DPP. This fast DPP method also inspires many follow-ups to improve diversity in different recommendation tasks~\cite{gan2020enhancing,liu2020diversified}.
	Meanwhile, SSD~\cite{huang2021sliding} proposed a time series analysis technique to include out-of-window items into the measurement of diversity to increase the diversity of a long recommendation sequence and alleviate the long tail effect as well. 
	% The recent FDSB~\cite{lin2022feature} proposed a disentangled attention encoder to extract item features and combine it with MMR to increase diversity in relevance recommendation. 

	\smallskip\noindent\textbf{User Interest Modeling.} 
	Researchers are capturing shifting user interests by modeling behavior sequences. For example, DIN~\cite{zhou2018deep} uses TA to capture user diversity, DIEN~\cite{zhou2019deep} uses GRU for drifting temporal interest, and MIND~\cite{li2019multi} uses multi-vectors for dynamic interests. These models focus on short sequences (<100). For long sequences, memory-based methods such as HPMN~\cite{ren2019lifelong} and MIMN~\cite{pi2019practice} use memory networks to model diverse user interests, while two-stage methods such as SIM~\cite{qi2020search} and UBR4CTR~\cite{qin2020user} train retrieval and CTR models separately. 
 	In the first stage, the retrieval model retrieves the top-$k$ relevant items from long user behavior sequences and stores the subsequence in an offline database.
	Then, in the second stage, the CTR model retrieves the top-$k$ relevant items directly from the offline database to reduce complexity during the learning. 
	These models mainly focus on the CTR tasks with the goal of maximizing accuracy. 
	Their successes in CTR prediction inspire us to extract user interests from both the long and short behavior sequences.
 
 % The retrieval model retrieves top-$k$ relevant items and stores them in a database, then the CTR model retrieves from the database to reduce complexity. These models focus on maximizing CTR accuracy. Extracting user interests from both long and short sequences is an inspiration.
	
	\section{Conclusion}
	In this paper, we propose a general re-ranking framework named MPAD for practical feed recommendation. A series of collaborative models are proposed to sequentially evaluate the accuracy and diversity of different items in a list and to generate an optimal item list by maximizing the joint utility of accuracy and diversity of the entire list. Both online and offline experiments verified the effectiveness of the proposed framework.
	
	\bibliographystyle{plain}
	\bibliography{DiversifiedRerank}
	%          Section
	% ============================
%	\clearpage
%	\newpage
	\appendix
	\section{Derivation of Item Selection}
%	\subsection{Definition of DPP}
%	A point process $ \mathcal{P} $ defined on a set $ I = \{1,2,\cdots,n\} $ is a probability distribution on the powerset of $ I $~(i.e., the set of all subsets of $ I $), where the probability satisfies $ \sum_{S \subseteq I}  \mathcal{P}(S) = 1 $.
%	For every subset $ S \subseteq I $, the probability of choosing $ S $ can be written as 
%	\begin{equation}\label{key}
%		\mathcal{P}(S) = \frac{\det\left(\boldsymbol{K}^u_{S}\right)}{\det\left(\boldsymbol{K}^u_I + \boldsymbol{E}\right)},
%	\end{equation}
%	where $ \det(\cdot) $ denotes the determinant of a matrix, $ \boldsymbol{K}^u_I \in \mathbb{R}^{n \times n} $ is a real positive semidefinite~(PSD) kernel matrix indexed by the elements of $ I $, $ \boldsymbol{K}^u_{S} \in \mathbb{R}^{k \times k}  $ denotes the restriction of $ \boldsymbol{K}^u_I $ to the entries indexed by the elements of $ S $, and $ \boldsymbol{E} $ is an identity matrix.
%	
%	The elementary kernels such as $\boldsymbol{D}^u_{\text{long}}$ and $\boldsymbol{D}^u_{\text{short}}$ are SE kernels constructed based on the learned interest vectors. 
	The composite kernel ${\boldsymbol{D}^u_S}$ is a PSD matrix since it is an addition of multiple PSD elementary kernels.
%	a PSD matrix for , such that all of its principal minors are also PSD.
	The Cholesky decomposition of ${\boldsymbol{D}^u_S}$ can be written as ${\boldsymbol{D}^u_S}={\bf V}{\bf V}^\top$, where ${\bf V} \in \mathbb{R}^{k \times k} $ is an invertible lower triangular matrix.
	For any $i \in { I \setminus S }$, the Cholesky decomposition of $\boldsymbol{D}^u_{S\cup\{i\}}$ can be represented as
	\begin{equation}\label{cho:dec:i}
		\boldsymbol{D}^u_{S\cup\{i\}}=
		\begin{bmatrix}
			\boldsymbol{D}^u_S & \boldsymbol{D}^u_{S,i} \\
			\boldsymbol{D}^u_{i,S} & \boldsymbol{D}^u_{ii}
		\end{bmatrix}=
		\begin{bmatrix}
			{\bf V} & {\bf 0} \\
			{\bf c}_i & d_i
		\end{bmatrix}
		\begin{bmatrix}
			{\bf V} & {\bf 0} \\
			{\bf c}_i & d_i
		\end{bmatrix}^{\top},
	\end{equation}
	where the row vector ${\bf c}_i$ and the scalar $d_i\ge0$ satisfies
	\begin{subequations}
		\begin{align}
			{\bf V}{\bf c}_i^{\top}&=\boldsymbol{D}^u_{S,i},\label{ci:direct} \\
			d_i^2&=\boldsymbol{D}^u_{ii}-\norm{{\bf c}_i}_2^2.\label{di:direct}
		\end{align}
	\end{subequations}
	According to~\eqref{cho:dec:i}, we have 
	\begin{equation}\label{di:intep}
		\det\left(\boldsymbol{D^u}_{S\cup\{i\}}\right)=\det\left({\bf V}{\bf V}^{\top}\right)\cdot d_i^2=\det\left(\boldsymbol{D}^u_{S}\right)\cdot d_i^2.
	\end{equation}
	Combine~(\ref{opt1-2}) with~(\ref{di:intep}), we obtain
	\begin{align}\label{opt3}
		j = \arg\max_{i \in I \setminus S} g(u,i|S) + \alpha \cdot \log (d^2_i).
	\end{align}
	We follow~\cite{chen2018fast} to derive the update of $ \log (d^2_i) $ as follows.
	The Cholesky decomposition of $\boldsymbol{D}^u_{S\cup\{j\}}$ can be written as
	\begin{equation}\label{chole:decom}
		\boldsymbol{D}^u_{S\cup\{j\}}=
		\begin{bmatrix}
			{\bf V} & {\bf 0} \\
			{\bf c}_j & d_j
		\end{bmatrix}
		\begin{bmatrix}
			{\bf V} & {\bf 0} \\
			{\bf c}_j & d_j
		\end{bmatrix}^{\top}.
	\end{equation}
	Define ${\bf c}'_i$ and ${d}'_i$ as the new vector and scalar of $i\in I\setminus(S\cup\{j\})$ after adding item $ j $ into $ S $.
	According to~\eqref{ci:direct} and~\eqref{chole:decom}, we have
	\begin{equation}\label{ci:equation}
		\begin{bmatrix}
			{\bf V} & {\bf 0} \\
			{\bf c}_j & d_j
		\end{bmatrix}
		{{\bf c}'_i}^{\top}
		=\boldsymbol{D}^u_{S\cup\{j\},i}
		=\begin{bmatrix}
			\boldsymbol{D}_{S,i} \\
			\boldsymbol{D}_{ji}
		\end{bmatrix}.
	\end{equation}
	Combining~\eqref{ci:equation} with Eq.~\eqref{ci:direct}, we have
	\begin{equation}\label{eq:ciUpdate}
		{\bf c}'_i=
		\begin{bmatrix}
		{\bf c}_i & (\boldsymbol{D}^u_{ji}-\langle{\bf c}_j,{\bf c}_i\rangle)/d_j
		\end{bmatrix}
		\doteq \begin{bmatrix}{\bf c}_i & e_i\end{bmatrix}.
	\end{equation}
	Then~\eqref{di:direct} implies
	\begin{equation}\label{di:update}
		d_i^{\prime2}=\boldsymbol{D}^u_{ii}-\|{\bf c}'_i\|_2^2=\boldsymbol{D}^u_{ii}-\|{\bf c}_i\|_2^2-e_i^2=d_i^2-e_i^2.
	\end{equation}
\end{document}